# Design Rules and Discovery of Face-Sharing Hexagonal Perovskites

M. J. Swamynadhan,[†,*] Gwan Yeong Jung,[†] Pravan Omprakash,[‡] & Rohan Mishra[†,‡,*]

[†]*Department of Mechanical Engineering and Material Science, Washington University in St. Louis, St. Louis, MO, 63130, United States*

[‡]*Institute of Materials Science and Engineering, Washington University in St. Louis, St. Louis, MO 63130, USA*

E-mail: swamynadhan@wustl.edu; rmishra@wustl.edu

**Abstract**

Hexagonal perovskites with face-sharing octahedral connectivity are an underexplored class of materials. We propose quantitative design principles for stabilizing face-sharing $ABX_3$ hexagonal perovskites based on a comparative analysis of oxides and sulfides. By mapping structural preferences across a phase-space defined by an electronegativity-corrected tolerance factor and the Shannon radius of the *A*-site cations, we identify distinct thresholds that separate hexagonal phases from competing cubic polymorphs having corner-sharing octahedral connectivity. Our analysis reveals that sulfides differ significantly from oxides due to the increased covalency of the transition metal–sulfur bonds, which enables broader compositional flexibility. Applying these principles, we predict a set of thermodynamically formable $AB\mathrm{O}_3$ and $AB\mathrm{S}_3$ compounds that are likely to adopt face-sharing octahedral connectivity. These findings establish a predictive framework for designing hexagonal perovskites, highlighting sulfides as promising candidates for obtaining quasi-one-dimensional materials having transition-metal cations for novel ferroic phenomena.

## Introduction

Inorganic perovskites with the composition $ABX_3$, where $A$ and $B$ are cations with different ionic radii, and $X$ is an anion, show a diverse range of physical properties. This versatility stems from the structural flexibility of the perovskite lattice to accommodate a wide range of cations. In response to the differing ionic radii, the ideal, aristotype cubic perovskite structure can undergo different distortions and connectivity changes to form various polymorphs, including orthorhombic perovskites ('Ortho')[1,2], post-perovskites ('Post')[3], face-sharing hexagonal perovskites ('$nH$')[4], and non-perovskite structures such as hexagonal manganites ('$h$−$R$MnO$_3$')[5] and ilmenites[6], as shown in Figure 1a. The various polymorphs of $ABX_3$ perovskites are classified based on the connectivity and arrangement of the $BX_6$ octahedra into corner-, edge-, or face-sharing geometries. While *"perovskite"* technically refers to corner-sharing $BX_6$ frameworks, the term "*hexagonal perovskite"* commonly denotes $ABX_3$ compounds having face-sharing octahedra or mixed connectivity (e.g., $BaNiO_3$, $BaMnO_3$, $BaTiS_3$)[7–10]. We adopt this convention for consistency with previous studies[11]. These structural variations have a strong impact on their electronic, magnetic, and optical properties, and enable a wide range of functionalities[11].

The structural flexibility of perovskites also makes them ripe for using materials informatics, often combined with density-functional theory calculations, for high-throughput screening and discovery of new compositions with properties tailored for targeted applications[12–15]. A key descriptor for the screening of stable $ABX_3$ polymorphs is the geometry-based Goldschmidt tolerance factor[16], $\tau$, which is defined as:

$$\tau = \frac{R_A + R_X}{\sqrt{2}(R_B + R_X)}, \quad (1)$$

where $R_A$, $R_B$, and $R_X$ are the ionic radii of the $A$−cation, $B$−cation, and $X$−anion, respectively. The above relationship comes from the ability to fit hard spheres into the perovskite framework. Specifically, it is a measure of the fit of the $A$-cation within the cavity formed by the corner-connected $BX_6$ network. Therefore, $\tau$ is suitable for screening corner-connected perovskite polymorphs, with $\tau = 1$ corresponding to the cubic structure, and $\tau < 1$ resulting in octahedral rotations and tilts that lower the symmetry to either the orthorhombic or the rhombohedral phase[17]. $\tau$ has been especially effective for screening oxides perovskites[1,18,19]. The different corner-connected structural variants of oxide perovskites can be effectively mapped in the $\tau - R_A$ phase space, as shown in Figure 1b, where each point represents a compound from the Materials Project database that lies on the convex hull[20]. Thus, $\tau$ and $R_A$ have been used as descriptors to screen and realize new oxide perovskites[21,22]. With the availability of large computational datasets and use of material informatics, modified versions of $\tau$ have been introduced to screen halide and chalcogenide perovskites[9,23] — that have attracted widespread interest in the past two decades for optoelectronic and solar-cell applications[24,25].

Generally in oxides, when the *A*-site cation is too large to fit within the ideal cubic perovskite framework, i.e., $\tau > 1$, the structure distorts into a hexagonal phase, positioning the *A*-cations between columns of face-sharing or edge-sharing octahedral chains[26]. The hexagonal polymorphs span the full spectrum—from the 100 % face-sharing 2*H* structure, to corner-sharing cubic phases[27,28]. The change in polyhedral connectivity leads to significant changes in the electronic structure, and results in unique properties that are not observed in the corner-connected counterparts[11]. A representative example is $BaMnO_3$, which exhibits multiple experimentally observed polytypes, including 33 % (6*H*), 50 % (4*H*) and 100 % (2*H*) face-sharing connectivity depending on the synthesis conditions[26]. The band gap increases by over 200 % as the percentage of face-sharing rises from 0 % (cubic) to 100 % (hexagonal)[10].

Beyond these structural distortions and variations in band gap, the introduction of face-sharing connectivity profoundly alters the local bonding environment and electronic dimensionality, giving rise to several emergent phenomena. Short *B-B* distances between magnetic ions in face-sharing octahedra form clusters arranged on a triangular lattice, producing competing intra- and inter-cluster interactions that lead to geometric and exchange frustration and can stabilize spin-liquid-like ground states, as reported in trimer-based oxides such as $Ba_4Nb\mathit{TM}_3O_{12}$ (*TM* = Mn, Ru, Ir) compounds [29,30]. In compounds such as $Sr_{9/8}TiS_3$, the one-dimensional chain-like electronic pockets give rise to colossal optical birefringence[31,32]. Strong electron–phonon coupling and non-negligible electron–electron interactions within these short *B–B* chains can drive charge-density-wave (CDW) and related electronic phase transitions, as observed in $BaTiS_3$ and related quasi-one-dimensional structures [33,34]. Strain-induced symmetry breaking in face-sharing frameworks couples axial octahedral rotations with ferroelectric polarization, enabling room-temperature chiral multiferroicity, recently demonstrated in $BaTiS_3$[35,36]. These emerging phenomena underscore the rich functional landscape uniquely accessible through face-sharing $ABX_3$ architectures.

Despite the rising interest in hexagonal perovskites, they remain comparatively underexplored. A search of hexagonal perovskites in the Inorganic Crystal Structure Database (ICSD)[37] results in only 78 $ABO_3$ and 10 $ABS_3$ compounds (and derivatives) having face-sharing octahedral chains (see Supplementary Information (SI) section 1 for the full list). In contrast, ICSD lists more than 5000 $ABO_3$ and around 250 $ABS_3$ compounds and their derivatives with the corner-connected perovskite framework. Furthermore, $\tau$ and $R_A$ cannot be used as descriptors for screening even hexagonal oxide perovskites, as their ranges coincide with those of corner-sharing cubic phases, as can be seen in Figure 1b. Thus, there is a need for quantitative design rules that can expand both oxide and sulfide-based face-sharing hexagonal perovskites.

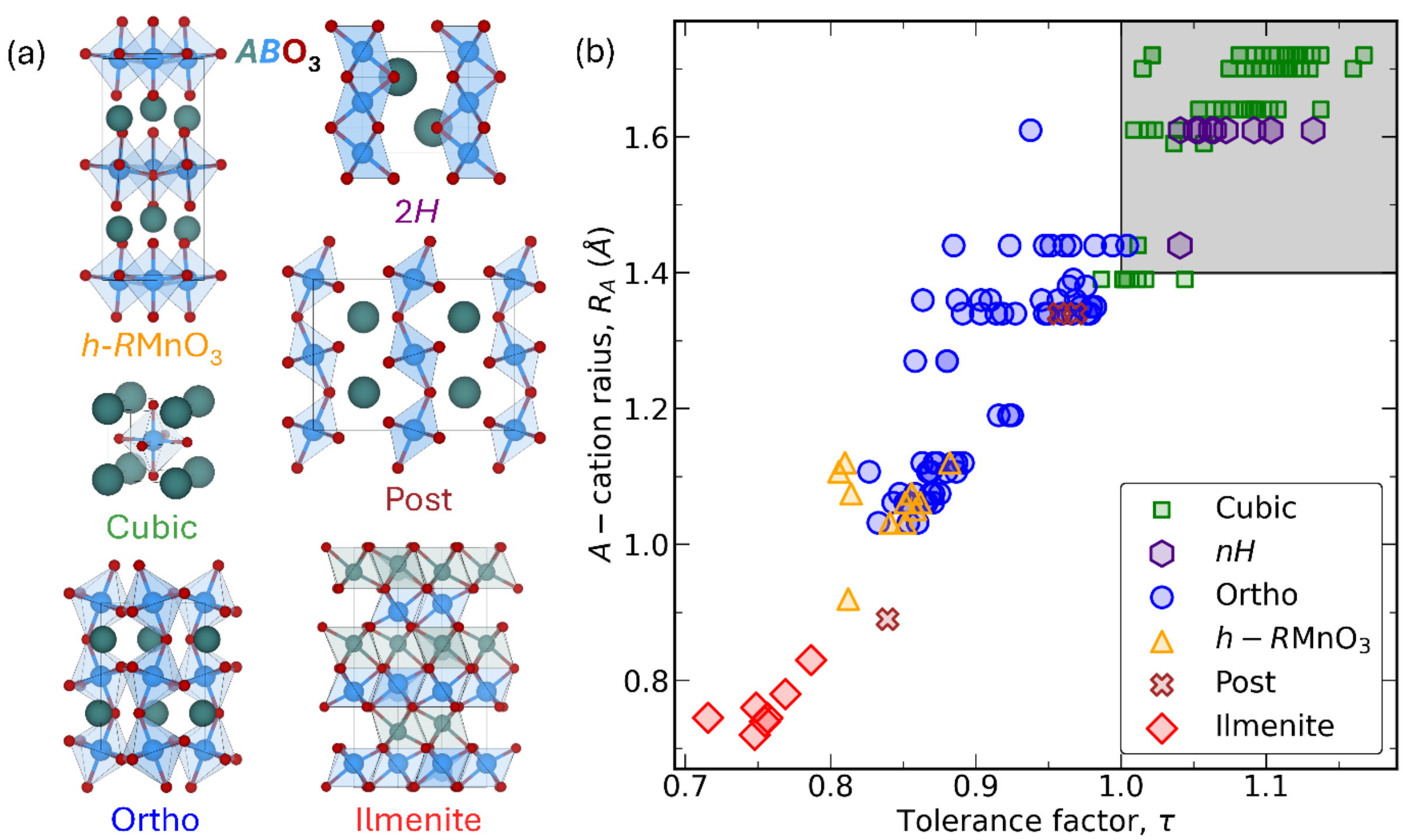

Figure 1: Structural variants of $ABO_3$. (a) Different structures with $ABO_3$ compositions. (b) Classification of different $ABO_3$ structures with respect to Goldschmidt tolerance factor, $\tau$ and $A$−cation radii, $R_A$, based on compounds on the energy hull obtained from the Materials Project database. The grey shaded region marks the range where the hexagonal perovskites appear.

In this Article, we present a framework for identifying $ABX_3$ compounds that are likely to form the face-sharing hexagonal perovskite phase and use it to predict 29 new $ABO_3$ and $ABS_3$ hexagonal perovskites. We used density functional theory (DFT) to calculate the formation energy of about 134 $ABO_3$ and 323 $ABS_3$ existing and hypothetical compositions in the face-sharing hexagonal and competing phases. Using this computed dataset, we mapped the stability of face-sharing hexagonal perovskite phases across a space defined by the tolerance factor, ionic radii, and the oxidation state of the cations. From the resulting trends, we derive design principles that highlight the geometric and electronic factors favoring face-sharing octahedral connectivity over conventional corner-sharing cubic structures. Applying these principles, we predict a set of $ABO_3$ and $ABS_3$ compounds likely to adopt the face-sharing hexagonal geometry. Notably, our analysis shows that sulfides offer greater structural flexibility, driven by a broader range of tolerance factors and more covalent *TM*–S bonding. This flexibility supports tuning strategies, such as cation substitution and double-perovskite-like arrangements, to design new hexagonal perovskite phases with targeted electronic, optical, and magnetic properties.

**Computational Details**

*DFT calculations:* All DFT calculations were performed using projector augmented-wave (PAW)[38] potentials as implemented in the Vienna Ab initio Simulation Package (VASP)[39]. The Perdew-Burke-Ernzerhof (PBE)-based generalized gradient approximation (GGA)[40] was used to treat exchange-correlation interactions with a plane-wave energy cutoff set at 520 eV. The Brillouin zone was sampled using a Γ-centered *k*-point mesh with a spacing of 0.025 $Å^{-1}$. Electronic self-consistent field calculations were considered converged when the total energy difference between successive iterations was less than $1 \times 10^{-8}$ eV. Both atomic positions and unit cell parameters were fully relaxed until the forces on each atom are less than $1 \times 10^{-3}$ eV/Å.

*Tolerance factor:* As discussed before, $\tau$ is effective for identifying broad structural trends in corner-connected oxide perovskites. However, in the case of sulfides, the classical $\tau$, defined above in Eq. 1, often fails to predict the phase stability of experimentally observed structures of many $ABS_3$ perovskites[41]. To address this limitation, we implemented two key corrections for sulfur-based systems:

1. *Updated ionic radii:* We used the revised Shannon radii[42] that is specifically adjusted for sulfur bonding environments.
2. *Modified tolerance factor ($\tau^*$):* We used a modified tolerance factor, $\tau^*$, proposed by Jess et al[9]., that takes into account the difference in the electronegativity of the cations and the anion. In this approach, the cation–anion bond lengths in the Goldschmidt tolerance factor are scaled by the relative change in the electronegativity difference $\Delta\chi$ compared to that of oxygen, as shown below:

$$\tau^* = \frac{\frac{\Delta\chi_{A-X}}{\Delta\chi_{A-O}}(R_A + R_X)}{\sqrt{2}\frac{\Delta\chi_{B-X}}{\Delta\chi_{B-O}}(R_B + R_X)}. \quad (2)$$

These corrections make phase predictions in corner shared sulfide perovskites more accurate and help better evaluate the stability of $ABS_3$ compounds[9,41].

<u>*Electrostatic force calculations*</u>*:* To evaluate long-range electrostatic interactions between *TM–TM* and *TM–X* pairs, we calculated the Coulombic force, $F$

$$F = \frac{M_{chain} q_i q_j}{4\pi\epsilon_0 R_{ij}}, \quad (3)$$

where, $q_i$ and $q_j$ are the partial charges, taken as the calculated average Bader charges of the $i^{th}$ and $j^{th}$ ions, respectively. $R_{ij}$ represents the distance between these ions in a fully relaxed structure. $M_{chain} = 2\ln 2 \approx 1.386$ is the one-dimensional Madelung constant for the hexagonal motif that accounts for all the periodic interactions in the infinite chain.

<u>*Selection of elements across the perovskite chemical space*</u>*:* The dataset of $ABO_3$ and $ABS_3$ compounds used in this study was systematically generated by combining selected elements from the Periodic Table that have been highlighted in Figure S1. *A*-site cations (shaded in green) include alkali metals, alkaline-earth metals, post-transition metals, and rare-earth elements. *B*-site cations (shaded in blue) consist of transition metals along with selected post-transition and rare-earth elements. Elements that can serve as both *A*-site and *B*-site cations are indicated by dual-color shading (green-blue) in Figure S1. Guided by the known stable oxidation states and coordination preferences, this approach produced 790 hypothetical $ABO_3$ and 790 hypothetical $ABS_3$ compositions that satisfy stoichiometry.

<u>*Lowest-energy polymorph:*</u> As discussed above, hexagonal oxides and sulfides exhibit a variety of crystal structures with different ratios of face-, edge- to corner-sharing octahedra. To identify the lowest-energy polymorph of each compound, we first limited our search to compositions with $\tau > 1$ — a regime where face-sharing hexagonal polytypes are energetically viable. Within this space, we constructed four representative polymorphs: corner-sharing cubic/orthorhombic perovskite, and face-sharing 2*H*, 4*H*, and 6*H* stackings (Figure S2). For the sulfides, we also included the fully edge-sharing, needle-like phase. We performed DFT relaxations for each candidate in all the considered crystal structures. The resulting total energies were compared to determine the lowest-energy polymorph. For magnetic compounds, we then re-examined the lowest-energy structure, starting from the ferromagnetic (FM) configuration and evaluating all symmetry-distinct collinear spin arrangements (see Figure S3) to identify the combined structural and magnetic ground state. Figure S4 presents the energy ordering across the four structural motifs. We then constructed energy-hull diagrams using the final energies—corresponding to the most stable polymorph with optimal spin alignment to quantitatively evaluate their formability and structural stability. These diagrams were generated by plotting the calculated formation energies of various competing phases, using phase-diagram analysis tools provided in the Materials Project database and pymatgen[43].

Note: All $\Delta E_{\text{hull}}$ values were obtained for the high-symmetry $P6_3/mmc$ parent structure. Previous first-principles studies on $BaTiS_3$, $CsTaS_3$, and related compounds show that this parent structure typically condenses one or more distortion modes, lowering the total energy by 8–22 meV atom$^{-1}$ while preserving face-sharing connectivity[32,35]. Thus, the stability map in Figure S4 and Table 1 may be interpreted as a conservative upper bound as the symmetry-lowering distortions are expected to further stabilize these phases.

## Results and Discussion

### Expansion and classification of the hexagonal $ABX_3$ chemical space

$ABX_3$ Hexagonal perovskites are rare, unlike their orthorhombic or cubic counterparts, with only a few examples reported in experimental or theoretical studies (see SI Tables S1 and S2). The scarcity of data makes it difficult to classify these hexagonal phases or define clear design principles. To address this issue, we systematically expanded the chemical space of plausible $ABX_3$ compositions, to identify candidates likely to adopt a face-sharing hexagonal motif. Details of element selection and dataset construction are provided in the Computational Details. In our classification scheme, we used Shannon's original ionic radii[44] for oxides and updated radii[42] for sulfides. Using these radii, we computed the modified tolerance factor, $\tau*$ (Eq. 2), for both the materials classes. Since $\tau*$ reduces to the classical $\tau$ in oxides, we adopt $\tau*$ as a unified descriptor for both oxides and sulfides throughout this work.

Our expanded chemical space contains 790 hypothetical $ABO_3$ and 790 hypothetical $ABS_3$ compositions that meet the 1:1:3 stoichiometry. Because we focus on face-sharing hexagonal phases, we restricted the search to compositions with $\tau^* > 1$. We obtain 134 oxides that satisfy these conditions. In sulfides, the electronegativity term in $\tau^*$ often pushes the value to double-digit numbers that, in practice, never crystallize in any perovskite-derived lattice. To avoid such unrealistic cases, we impose an upper bound of two and obtain 323 sulfide compositions that fall within $1 \leq \tau^* \leq 2$.

To classify these candidates accurately, previous studies have emphasized the need for separate classification of each valence family, as the charges on the cations can significantly alter the size tolerance[45,46]. We therefore classify $A^{1+}B^{5+}X_3$, $A^{2+}B^{4+}X_3$, and $A^{3+}B^{3+}X_3$ as separate families. The distributions of $R_A$ and $\tau^*$ for $A^{3+}B^{3+}$, $A^{2+}B^{4+}$, and $A^{1+}B^{5+}$ compounds are shown in SI (Section 3) Figure S5.

### $A^{3+}B^{3+}X_3$ family

We first examine the $A^{3+}B^{3+}X_3$ family of compounds. In oxides, the *A*-site radii range from 0.82 to 1.38 Å, placing most materials in the $\tau^* < 1$ region, as shown in Figures S5a and S5b. These smaller radii prevent the stabilization of face-sharing octahedral connectivity, constraining these compounds to either orthorhombic or hexagonal manganite structures with trigonal bipyramids such as the *h*-$RMnO_3$ type compounds (shown in Figure 1a). In sulfides, the correction factor ($\Delta\chi$) and the larger sulfur-based ionic radii of the cations shift many $A^{3+}B^{3+}S_3$ compounds into the $\tau^* > 1$ region, as clearly visible in Figure S5b. Yet, no member of this group adopts a face-sharing motif; instead, the structures settle into the fully edge-sharing "needle-like" chains seen in $NH_4CdCl_3$ or into the same orthorhombic and hexagonal manganite frameworks preferred by the oxides (shown in Figure S6). These observations underscore the critical influence of $R_A$ on structural preference. For $A^{3+}B^{3+}X_3$ compounds, smaller $R_A$ restricts the stabilization of face-sharing hexagonal phases, regardless of the $\tau^*$ or the anion involved.

### $A^{2+}B^{4+}X_3$ family

Unlike $A^{3+}B^{3+}X_3$ compounds, several members of the $A^{2+}B^{4+}X_3$ family fall into the $1 < \tau^* < 2$ region for both oxides (59) and sulfides (87) (Figure S5a and S5b) due to the larger radii of the $A^{2+}$ cations. To evaluate their structural preferences, we predicted the ground-state structure by comparing the energetics of common competing polymorphs for each $ABX_3$ compound with $1 < \tau^* < 2$ (see the Computational Details section for more details).

The phase distributions are presented in the $\tau^* - R_A$ phase space, as shown in Figure 2a for oxides and Figure 2b for sulfides. Corner-sharing polymorphs are represented as green squares (cubic) and blue circles (orthorhombic). face-sharing Hexagonal "*n*H" phases are shown by indigo hexagonal symbols and sky blue circles for orthorhombic edge-sharing "needle" structures[47]. Hollow markers correspond to previously reported compounds, while solid markers represent new predictions from our calculations. Our findings agree with Carr et al.[48], who screened 81 $A^{2+}B^{2+}S_3$ compositions with DFT and reported that 77 % adopt the needle-like motif, attributing this preference over face-sharing to electrostatic interactions.

A sharp boundary separates corner-sharing and face-sharing hexagonal phases at $\tau^* = 1.04$ for oxides and $\tau^* = 1.07$ for sulfides. However, simple geometric descriptors cannot distinguish fully face-sharing (2H) structures from partially face-sharing (4H/6H) ones in oxides, nor from edge-sharing needle-like structures in sulfides, and experimental studies show that temperature[47], pressure[7], and defect concentration[49] can toggle among these motifs. Thus, a comprehensive classification lies beyond the scope of this study.

In sulfides, a second size constraint becomes important: *A*-site cations with radii smaller than 1.7 Å favor orthorhombic corner-sharing or needle-like frameworks even when $\tau^*$ exceeds the 1.07 threshold. For *A*-site cations with radii greater than 1.7 Å can stabilize face-sharing hexagonal phases, provided the $\tau^*$ criterion is also met. This dual requirement explains why no $A^{3+}B^{3+}X_3$ sulfides adopt a hexagonal structure. Although many sulfides exceed the $\tau^*$ threshold, all have $R_A$ below 1.7 Å. On the other hand, most $A^{2+}B^{4+}X_3$ sulfides meet both the conditions and readily form face-sharing hexagonal frameworks.

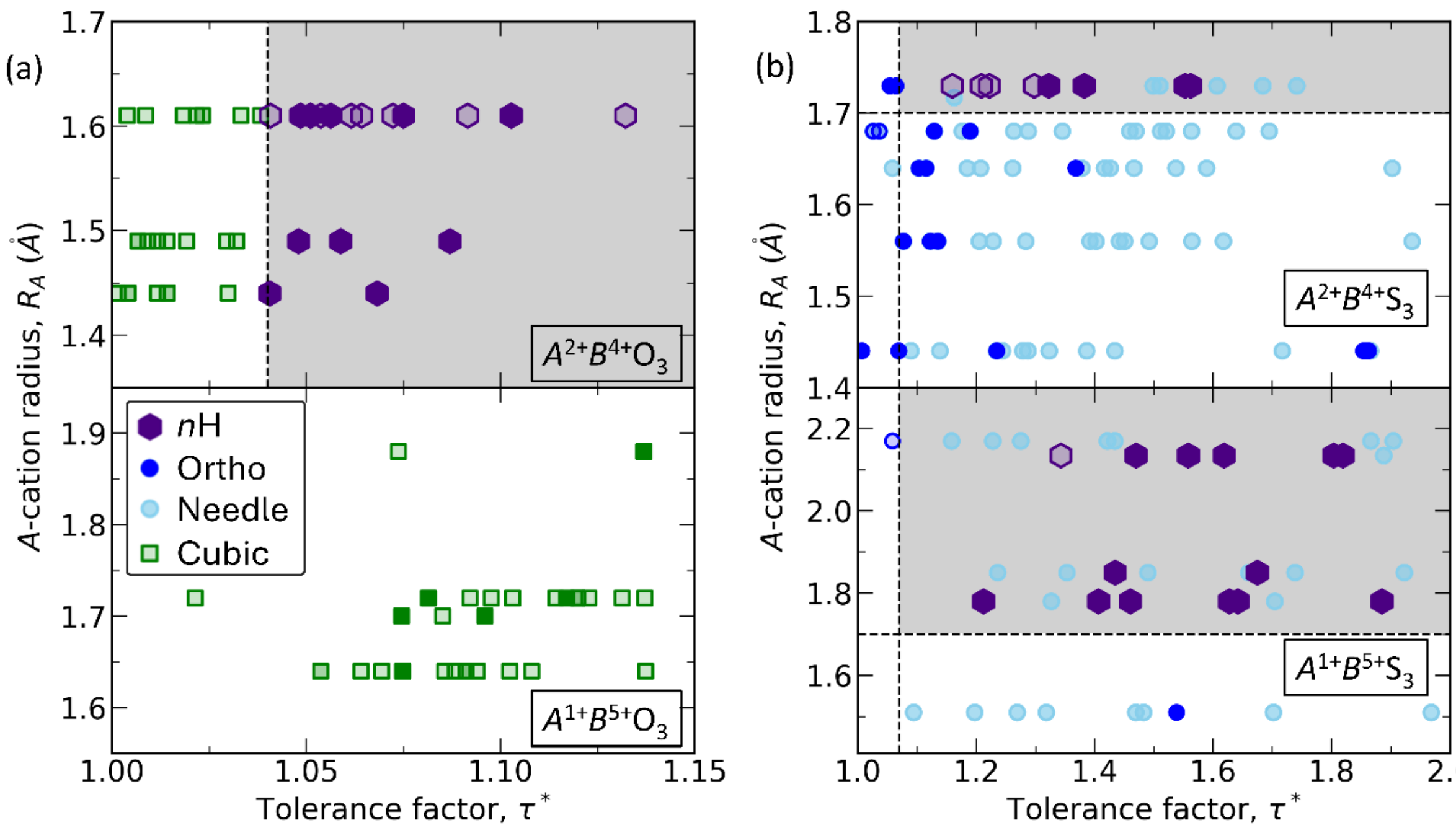

Figure 2: $\tau^*$ Vs $R_A$ scatter plots. (a) Classification of stable polymorphs in the phase space defined by $\tau^*$ and $R_A$ for $A^{2+}B^{4+}O_3$ (top panel) and $A^{1+}B^{5+}O_3$ (bottom panel). (b) Classification of competing $ABS_3$ polymorphs for both $A^{2+}B^{4+}S_3$ and $A^{1+}B^{5+}S_3$. Hollow markers represent compounds previously reported experimentally or theoretically, while solid markers indicate compounds predicted in this work. Some of these compositions have been experimentally realized in different polymorphs, as discussed in the section "Comparison with experimentally reported phases.

### $A^{1+}B^{5+}X_3$ family

We next investigate $A^{1+}B^{5+}X_3$ compounds. Among approximately 70 $A^{1+}B^{5+}O_3$ compositions with $\tau^* > 1$ and sufficiently large $R_A$ values (Figure S5a), none were found to stabilize in a face-sharing hexagonal arrangement—neither in our calculations nor in prior literature (Figure 2a). In contrast, of the 50 $A^{1+}B^{5+}S_3$ compositions with $1.0 < \tau^* < 2.0$, 14 stabilize into face-sharing hexagonal structures (Figure 2b). As in the $A^{2+}B^{4+}X_3$ family, a clear structural boundary at $\tau^* = 1.07$ separates corner-sharing orthorhombic phases from hexagonal ones and $R_A \geq 1.7$ Å is required to stabilize the face-sharing motif over competing edge-shared polymorphs. While $A^{1+}B^{5+}O_3$ favor only corner-sharing structures, sulfides consistently stabilize into face-sharing hexagonal structures, consistent with both reported data and our DFT-based predictions. A well-known example is $CsTaS_3$[50], an experimentally observed face-sharing hexagonal sulfide containing $Cs^{1+}$ and $Ta^{5+}$.

Overall, these contrasting outcomes raise the question as to why the face-sharing hexagonal structure is so selective toward specific cation charge states.

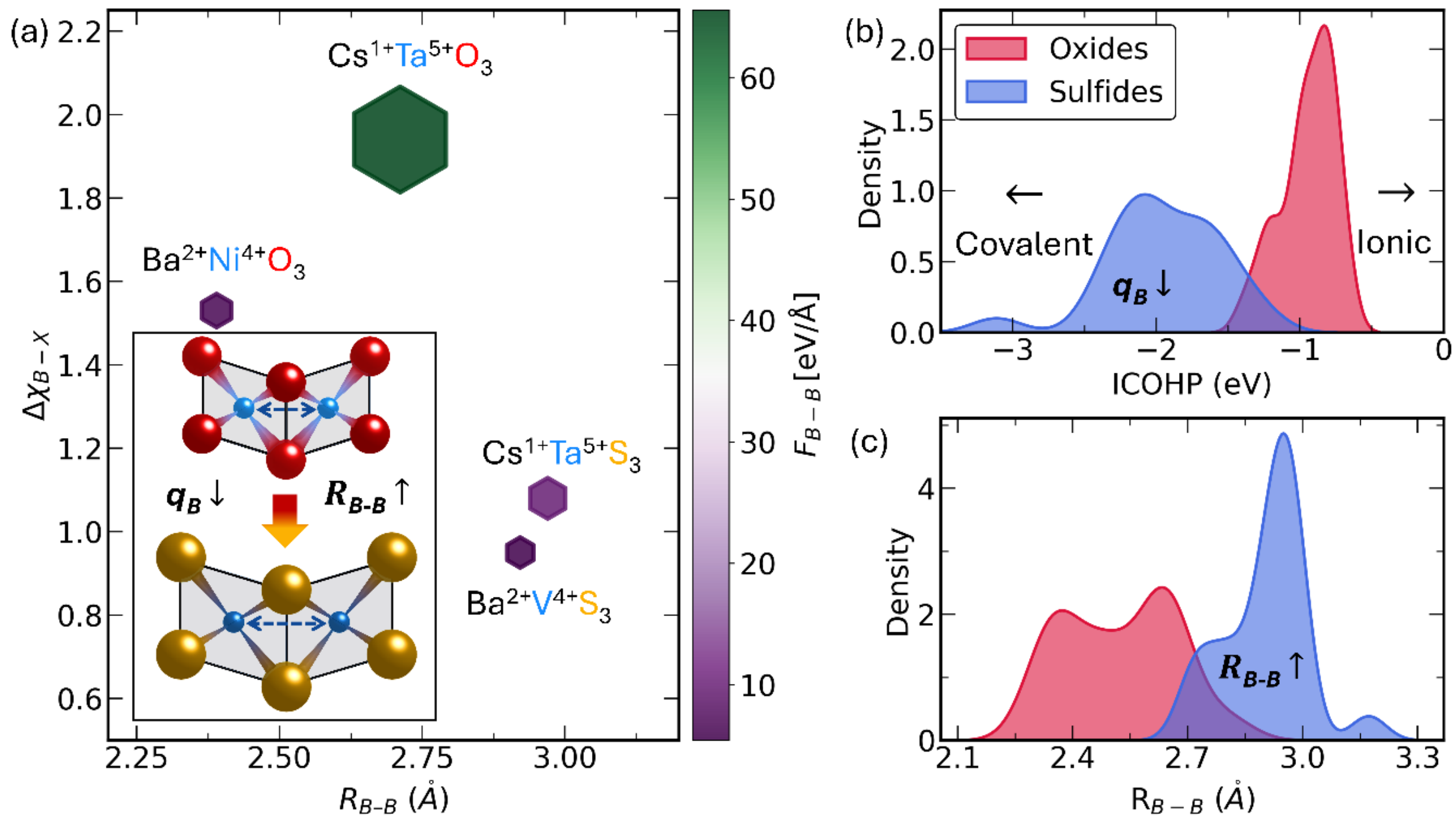

Figure 3: Charge state selectivity. (a) Coulombic force between face-sharing $B$–cations, $F_{B\text{–}B}$ as a function of $\Delta\chi_{B\text{–}X}$ and $R_{B\text{–}B}$. The size and color of the hexagonal markers represent the magnitude of $F_{B\text{–}B}$. (b) Distribution of integrated Crystal Orbital Hamilton Population (ICOHP) for oxide and sulfide bonds and (c) inter-ionic radius ($R_{B\text{–}B}$) calculated for oxides and sulfides.

**Selectivity towards oxidation state**

To investigate why oxides and sulfides behave differently at similar $\tau*$ and $R_A$, we examined four representative experimentally observed compounds having similar tolerance factor ($\tau^*$~1.3): $BaNiO_3$[51], $CsTaO_3$[52], $BaVS_3$[53], and $CsTaS_3$[50]. Among the oxides, $Ba^{2+}Ni^{4+}O_3$ adopts a face-sharing hexagonal ground state, whereas $Cs^{1+}Ta^{5+}O_3$ favors a corner-sharing cubic structure. Remarkably, both $Ba^{2+}V^{4+}S_3$ and the same $Cs^{1+}/Ta^{5+}$ pair stabilize in a face-sharing hexagonal lattice when sulfur replaces oxygen ($Cs^{1+}Ta^{5+}S_3$). As discussed in the previous section, none of the $A^{1+}B^{5+}O_3$ compounds we examined favor the face-sharing hexagonal polymorph. To understand this selectivity towards oxidation state, we calculated the Coulombic forces between neighboring $B$-cations (Eq. 3) in the face-sharing arrangement for the four compounds. The results are plotted in Figure 3a as a function of $\Delta\chi_{B\text{–}X}$ and $R_{B\text{–}B}$, where the size and the color of the hexagonal markers represent the magnitude of the Coulombic force $F_{B\text{–}B}$. The instability of hexagonal $Cs^{1+}Ta^{5+}O_3$ is attributed to the high Coulombic repulsion between neighboring *B* cations. In face-sharing octahedral networks, the *B*−*B* distances are shorter, and highly charged ions, such as $Ta^{5+}$, experience significant electrostatic repulsion that destabilizes the structure. In contrast, $Cs^{1+}Ta^{5+}S_3$ and $Ba^{2+}V^{4+}S_3$ experience substantially lower repulsive forces due to increased *B*–S covalency compared to *B*–O. This enhanced covalency has two key effects: (1) it increases the *B–B* distance from 2.71 Å in $CsTaO_3$ to 2.97 Å in $CsTaS_3$, and (2) the increased *B*–S covalency reduces the effective charge on the *B* cations. Together, these effects decrease the net Coulombic repulsion, stabilizing the face-sharing hexagonal phase in sulfides. The effect of the anions on the charge distribution and interionic distance is illustrated as an inset in Figure 3a. To quantify the charge distribution and covalency, we compared the integrated Crystal Orbital Hamilton Population[54] (ICOHP) values for the *B–X* bonds and interionic *B*−*B* distances ($R_{B\text{–}B}$) across all the 134 oxides and 137 sulfides considered in this study (Figure 3b). Negative ICOHP values indicate stronger covalency and coincide with longer *B–B* distances; both trends are most pronounced in sulfides. Having established that stronger *TM*–S covalency unlocks face sharing in $A^{1+}B^{5+}S_3$, we next explore how the broader distribution in τ* values in sulfides can be harnessed to design new stable hexagonal phases.

**Tuning τ* to stabilize face-sharing sulfides**

For oxides, $\tau*$ and $R_A$ provide a clear classification of face-sharing hexagonal phases. For sulfides, these geometric descriptors separate 3D corner-sharing from 1D chain motifs but do not clearly distinguish face-sharing from edge-sharing connectivity; geometric descriptors alone are insufficient. Nevertheless, in the Ba-based sulfides a systematic progression emerges as $\tau*$ increases. From $\tau^* \leq 1.07$, 3D corner-sharing perovskites with the *Pnma* space group are stable (e.g., $BaHfS_3$, $BaZrS_3$). For $1.07 < \tau^* < 1.4$, face-sharing hexagonal phases having 1D chains appear (e.g., $BaTiS_3$, $BaNbS_3$). For $\tau^* \geq 1.4$, “needle-like” structures with 1D edge-sharing chains become dominant. These phases are related to the $NH_4CdCl_3$ family, and $BaGeS_3$ in particular crystallize in $P2_1/c$ space group. These structures are shown in Figure 4a.

Guided by this trend, we propose a design strategy that leverages the broad span of τ* values of sulfides: tune τ* into the face-sharing stability window (1.07–1.4) using solid solutions. As a test case, we studied $BaHf_{(1-x)}Ge_xS_3$. Based on our DFT stability calculations, the end member $BaHfS_3$[55] ($\tau^* = 1.05$) adopts a corner-sharing orthorhombic perovskite structure (*Pnma*), while $BaGeS_3$[56] ($\tau^* = 1.74$) crystallizes in a distorted 1D edge-sharing phase ($P2_1/c$), and both the structures have been reported experimentally[55,56]. Neither end member is face-sharing or hexagonal. We generated ordered and special quasi-random[57] (disordered) models for $x = 0.25$, 0.50, and 0.75. (Figure S7); the disordered solutions are lower in energy than the ordered double-perovskite arrangements. We compared the energetics of the *Pnma*, $P6_3/mmc$ (face-sharing hexagonal), and $P2_1/c$ space groups. Wherever the composition-weighted τ* falls within the face-sharing window, $P6_3/mmc$ is either the ground state or within 10 meV per formula unit of the lowest phase. The composition-dependent energy comparison for these three motifs across $x = 0$ to 1 is shown in Figure 4b, with the stability regions shaded. This confirms that tuning τ* via solid solutions can stabilize face-sharing hexagonal networks, positioning sulfide perovskites as a platform for rational materials discovery.

**Design rules to obtain face-sharing hexagonal phases:**

Based on the above findings, we can distill the following rules for designing face-sharing hexagonal $ABX_3$ structures:

1. Select large *A*-cations to promote the cubic-to-hexagonal transition: $R_A \geq 1.4$ Å for oxides and 1.7 Å for sulfides.
2. Choose *B*-cations to tune $\tau^*$ into the stability window: $\tau^* > 1.04$ for oxides, and $1.07 < \tau^* < 2.0$ for sulfides.
3. Restrict the cation oxidation states appropriately: only $A^{2+}B^{4+}O_3$ oxides exhibit face-sharing hexagonal stability, while both $A^{2+}B^{4+}S_3$ and $A^{1+}B^{5+}S_3$ exhibit face-sharing hexagonal stability among sulfides.
4. Employ substitutional alloying or ordered double-perovskite structures in sulfides to tune $\tau^*$ into the optimal stability window for face-sharing stability.

Following these design rules, we have predicted a set of oxides and sulfides that favor the face-sharing hexagonal structure. Their preferred octahedral connectivity and energy above the convex hull ($\Delta E_{hull}$) are summarized in Table 1.

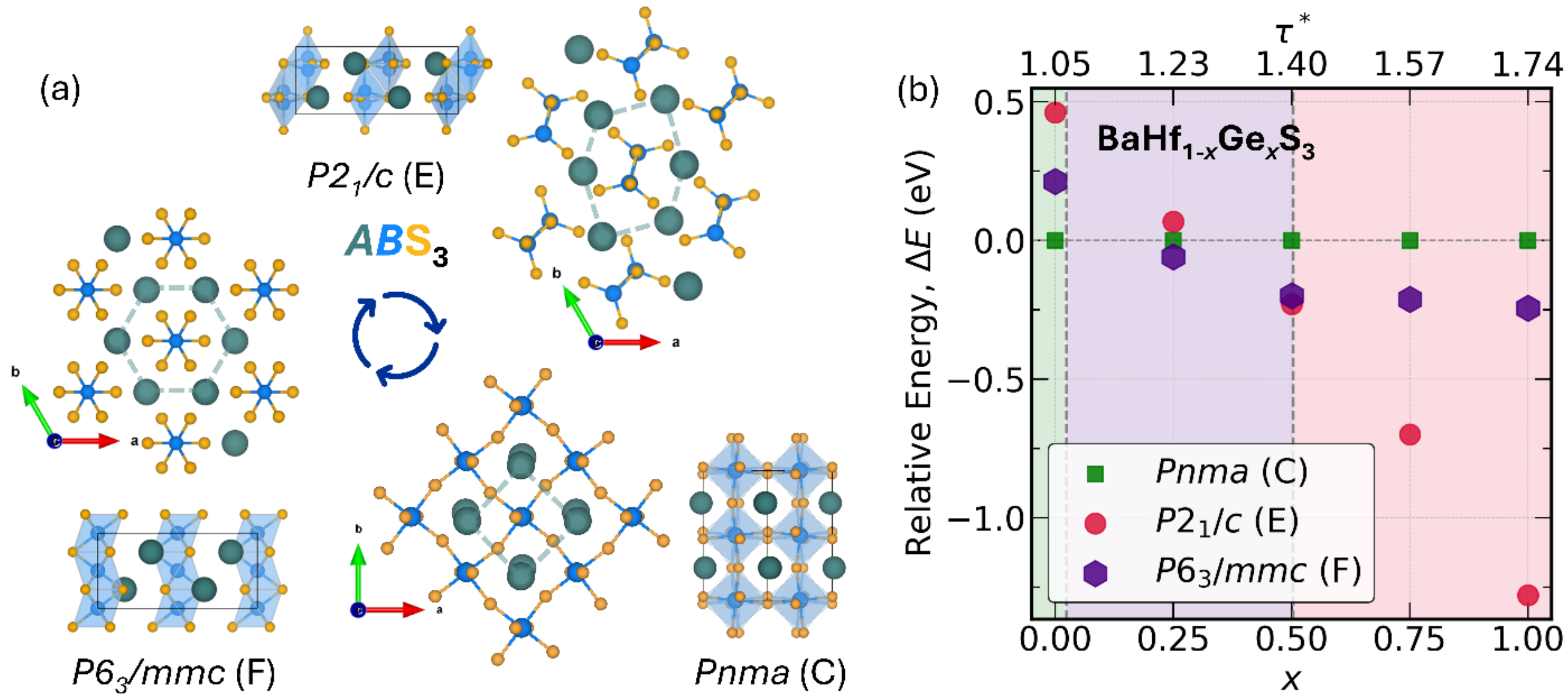

Figure 4. Tuning τ* to stabilize face-sharing sulfides. (a) Corner-sharing *Pnma,* (C) Face-sharing $P6_3/mmc$ (F) and edge-sharing $P2_1/c$ (E) phases in the $BaBS_3$ series. (b) Relative energies of corner-sharing (C), face-sharing (F), and edge-sharing (E) motifs for $BaHf_{(1-x)}Ge_xS_3$ solid solutions. Green squares denote corner-sharing *Pnma*, indigo hexagons denote face-sharing $P6_3/mmc$, and red circles denote edge-sharing $P2_1/c$; shaded regions indicate the stability ranges of the corresponding polymorphs.

**Comparison with experimentally reported phases:**

The most promising oxides and sulfides predicted to form hexagonal face-sharing phases are listed in Table 1. These compounds have not yet been reported in these hexagonal face-sharing phases experimentally but fall within the proposed $R_A - \tau^*$ stability regime. A detailed comparison between these predictions and experimentally realized compounds is discussed below. A more extensive list of systems predicted to form hexagonal face-sharing phases is provided in Table S2. Our predictions show broad agreement with experimentally known Ba- and Sr-based perovskites. Several $BaBO_3$ oxides ($B$ = Tc, Fe, Co, Ni, Mn, Ti, Ir, Pt, Rh, Cr)[58–67] and $SrMnO_3$[68], as well as the sulfides $BaMS_3$ ($M$ = Ta, Nb, Ti, V)[69–72] and $CsTaS_3$[50], are experimentally known to adopt face-sharing polytypes, consistent with our predictions. None of the sulfide compositions listed in Table 1 have yet been synthesized; however, all known face-sharing sulfides (e.g., $BaTiS_3$, $CaTaS_3$) fall within our predicted stability window, indicating that the newly proposed compositions are realistic candidates for synthesis.

Two notable outliers are $BaOsO_3$[73] and $SrCoO_3$[74]. Both are experimentally reported as cubic perovskites stabilized only under high-pressure, high-oxygen conditions and lie at $\Delta E_{hull} \approx 80\text{meV/atom}$[20], whereas our predicted face-sharing phases are more stable at $\Delta E_{hull} = 6\text{meV/atom}$ and 20 meV/atom, respectively. Similarly, $BaVO_3$[75] and $PbMnO_3$[76] are reported in $5H$ and $6H$ polytypes ($\Delta E_{hull} = 75$ and $72\text{meV/atom}$)[20], while our $6H$ and $4H$ models are lower in energy at $\Delta E_{hull} = 53$ and $62\text{meV/atom}$, respectively. These discrepancies between the predicted and the observed phases are likely due to synthesis conditions. Our model identifies face-sharing as the thermodynamically preferred connectivity for these compositions; however, experimental factors, such synthesis pressure, quenching rate, and oxygen partial pressure, can override the thermodynamic preference and drive the system into alternative motifs, as discussed by Nguyen and Cava (2021)[8]. This sensitivity to synthesis conditions is well established in Ba- and Sr-based perovskites, particularly $BaMnO_3$[26] and $BaTiO_3$[63,77,78], which exhibit multiple connectivity under different synthesis conditions.

For other oxides in Table 1, the experimentally known cubic or orthorhombic polymorphs typically lie at $\Delta E_{hull} > 100\ \text{meV/atom}$[20]. Although their predicted face-sharing phases lie within a comparable energy range to reported polymorphs, face-sharing phases in such compounds are experimentally accessible under favorable synthesis conditions.

Phases with $\Delta E_{\text{hull}} \leq 50$ meV/atom are generally considered formable, although the formability limit can be tightened based on the class of the compounds[79]. The $\Delta E_{\text{hull}}$ values reported herein are for the high-symmetry $P6_3/mmc$ phase because this structure uniquely captures the face-sharing motif that motivates our design rules. Previous first-principles studies on $BaTiS_3$[35], $CsTaS_3$,[33,36] and related compounds indicate that the parent $P6_3/mmc$ phase typically contains one or more *soft* phonon instabilities that reduce the symmetry by undergoing various distortions to lower the total energy by 8–22 meV/atom. Therefore, several of the formable compounds listed in Table 1 with the high-symmetry $P6_3/mmc$ phase can be expected to be more stable in a lower-symmetry sub-group structure, and the calculated $\Delta E_{\text{hull}}$ serves as a conservative upper bound.

**Conclusions**

We have formulated quantitative design rules for face-sharing hexagonal perovskites and applied them to predict new compounds in the $ABO_3$ and $ABS_3$ chemical space. These rules are based on the electronegativity-corrected tolerance factor ($\tau^*$) and the $A$-site cation radius ($R_A$). We have identified two practical windows (Figure 2):

- Oxides: $\tau^* > 1.04, R_A > 1.4$ Å
- Sulfides: $\tau^* > 1.07, R_A > 1.7$ Å

Although the oxide and sulfide windows are similar, the greater covalency of *B*–S bonds broadens the range of $\tau^*$ values possible and opens up alloying or forming solid solutions or (e.g., $BaHf_{(1-x)}Ge_xS_3$) to achieve compositions within the face-sharing stability window.

While this work establishes general design principles governing the emergence of face-sharing hexagonal frameworks, it does not yet distinguish the energetic preferences among the different hexagonal polytypes and mixed face-sharing motifs. To probe the origins of these preferences, we examined several structural and bonding descriptors, including bond-valence sums, anion packing, and *A*-site volume distortion. Although no single quantitative parameter could fully capture this trend within the current dataset, a qualitative picture emerges. In sulfides, the larger ionic radius and more covalent *B*-S bonding reduce electrostatic repulsion across shared faces and allow the lattice to accommodate shorter *B*–*B* separations, making the continuous, face-sharing octahedral chains of the 2*H* structure energetically favorable. In contrast, oxides contain the smaller and highly electronegative $O^{2-}$ anions; the stronger, more ionic *B*–O bonds and greater *A*-site mismatch promote the stability of polytypes with interrupted (shorter) face-sharing chains, such as 4*H* and 6*H*. A deeper quantitative understanding of these stacking variations arising from competing geometric, electrostatic, and covalent factors will require a dedicated study.

From a functional perspective, the predicted hexagonal compounds with $d^0$-cations, such as $CsVS_3$, $RbVS_3$ and $RbTaS_3$ are especially promising as they can show novel ferroic phenomena such as coupled ferroelectricity and chirality[35], non-collinear polar textures driven by flat-band instabilities[35], both of which have been recently reported in $BaTiS_3$. The quasi-1D structure of these compounds can also result in highly anisotropic physical properties including birefringence and anisotropic thermal transport[71,80]. Furthermore, face-sharing systems containing partially filled *d* shells, such as $BaOsO_3$, $BaVO_3$, $BaReS_3$, and $CsCrS_3$, provide fertile ground for exploring exotic magnetic interactions and possible spin-liquid behavior[29,30]. These structure–property relationships represent exciting directions for future investigation into the functionalities of the hexagonal $ABX_3$ compounds identified in this study.

**Acknowledgements**

The authors gratefully acknowledge Jayakanth Ravichandran and Shantanu Singh for helpful discussions. This work was primarily supported the National Science Foundation (NSF) through grant nos. DMR-2122070 (M.J.S., G.Y.J., R.M.) and DMR-2145797 (P.O., R.M.). This work used computational resources through allocation DMR160007 from the Advanced Cyberinfrastructure Coordination Ecosystem: Services & Support (ACCESS) program, which is supported by NSF grants #2138259, #2138286, #2138307, #2137603, and #2138296. The authors thank the anonymous reviewer for a careful reading of the manuscript and for several insightful suggestions that helped improve the clarity of the paper.

**Supporting Information**

ICSD survey of all known hexagonal $ABO_3$ and $ABS_3$ compounds; $R_A - \tau^*$ distributions; structural variants and spin configurations of *nH* polytypes; energy comparison of polymorphs; $BaHf_{1-x}Ge_xS_3$ solid-solution models and SQS structures; complete list of predicted face-sharing $ABX_3$ compounds with $R_A$, $\tau^*$, face-sharing percentage, and $\Delta E_{hull}$ tables.

**Data Availability Statement**

The data supporting this study are openly available in the repository[81] *"Design Rules and Discovery of Face-Sharing Hexagonal Perovskites"* at https://doi.org/10.5281/zenodo.17636148.

Table 1: Predicted face-sharing hexagonal compounds with respective $A$−cation radius $R_A$ (Å), tolerance factor ($\tau^*$), percentage of face-sharing, and energy above convex hull, $\Delta E_{\text{hull}}$ (meV/atom).

| No. | Compound | $R_A$ (Å) | $\tau^*$ | Face-Sharing (%) | $\Delta E_{\text{hull}}$ (meV/atom) |
|---|---|---|---|---|---|
| 1 | $BaOsO_3$ | 1.61 | 1.05 | 50 | 6 |
| 2 | $SrNiO_3$ | 1.44 | 1.07 | 100 | 7 |
| 3 | $SrCoO_3$ | 1.44 | 1.04 | 100 | 20 |
| 4 | $PbMnO_3$ | 1.49 | 1.06 | 50 | 53 |
| 5 | $BaVO_3$ | 1.61 | 1.07 | 33 | 62 |
| 6 | $BaReO_3$ | 1.61 | 1.05 | 50 | 85 |
| 7 | $BaGeO_3$ | 1.61 | 1.10 | 100 | 144 |
| 8 | $SrGeO_3$ | 1.44 | 1.04 | 33 | 154 |
| 9 | $PbCoO_3$ | 1.49 | 1.06 | 100 | 172 |
| 10 | $PbCrO_3$ | 1.49 | 1.05 | 33 | 181 |
| 11 | $PbGeO_3$ | 1.49 | 1.06 | 33 | 210 |
| 12 | $PbNiO_3$ | 1.49 | 1.09 | 100 | 480 |
| 13 | $BaMnS_3$ | 1.73 | 1.32 | 100 | 4 |
| 14 | $RbTaS_3$ | 1.78 | 1.21 | 100 | 8 |
| 15 | $CsNbS_3$ | 2.14 | 1.47 | 100 | 31 |
| 16 | $BaCrS_3$ | 1.73 | 1.38 | 100 | 56 |
| 17 | $CsVS_3$ | 2.14 | 1.56 | 100 | 90 |
| 18 | $RbVS_3$ | 1.78 | 1.41 | 100 | 101 |
| 19 | $RbSbS_3$ | 1.78 | 1.88 | 100 | 107 |
| 20 | $KVS_3$ | 1.85 | 1.43 | 100 | 114 |
| 21 | $CsCrS_3$ | 2.14 | 1.62 | 100 | 126 |
| 22 | $BaTcS_3$ | 1.73 | 1.55 | 100 | 129 |
| 23 | $CsReS_3$ | 2.14 | 1.82 | 100 | 130 |
| 24 | $RbCrS_3$ | 1.78 | 1.46 | 100 | 141 |
| 25 | $RbReS_3$ | 1.78 | 1.64 | 100 | 158 |
| 26 | $CsTcS_3$ | 2.14 | 1.80 | 100 | 169 |
| 27 | $KReS_3$ | 1.85 | 1.67 | 100 | 172 |
| 28 | $BaReS_3$ | 1.73 | 1.56 | 100 | 177 |
| 29 | $RbTcS_3$ | 1.78 | 1.63 | 100 | 185 |

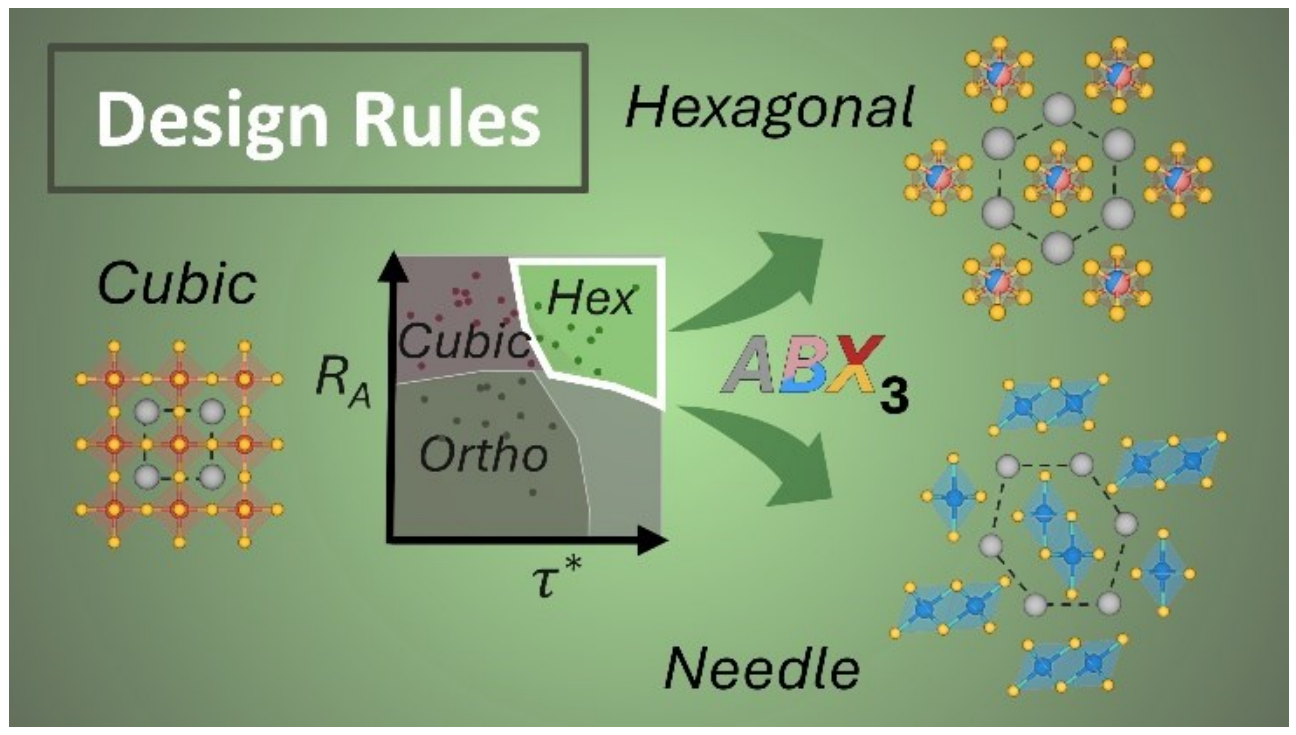

**Table of Contents Graphic**